\documentclass[%
 reprint,
 amsmath,amssymb,
 aps,
]{revtex4-2}

\usepackage{graphicx}
\usepackage{dcolumn}
\usepackage{bm}

\usepackage{amssymb}
\usepackage{array}
\usepackage{mathrsfs}
\usepackage{amscd}
\usepackage{amsmath}
\usepackage{amsfonts}
\usepackage{amsthm}
\usepackage{psfrag}
\usepackage{textcomp}
\usepackage{bm}
\usepackage{dsfont}
\usepackage{lineno}
\usepackage{color}
\usepackage{float}

\usepackage[dvipsnames]{xcolor}
\usepackage[colorinlistoftodos]{todonotes}
\setuptodonotes{fancyline, linecolor=BrickRed, backgroundcolor=BrickRed!10!white,
bordercolor=BrickRed, textcolor=black,inline}

\begin{document}

\preprint{APS/123-QED}

\title{
Fabry-Pérot quasinormal modes for topological edge states}

\author{Marc Mart\'i Sabat\'e}
 \affiliation{Department of Mathematics, Imperial College London, London SW7 2AZ, UK}
 \altaffiliation{m.marti-sabate23@imperial.ac.uk}
\author{Benjamin Vial}%
 \affiliation{Department of Mathematics, Imperial College London, London SW7 2AZ, UK}%
\author{Richard Wiltshaw}%
 \affiliation{Department of Mathematics, Imperial College London, London SW7 2AZ, UK}%

\author{S\'ebastien Guenneau}
\affiliation{The Blackett Laboratory, Department of Physics, Imperial College London, London SW7 2AZ, UK}%
\author{Richard V. Craster}
\affiliation{Department of Mathematics,
    UMI 2004 Abraham de Moivre-CNRS,
    Department of Mechanical Engineering, Imperial College London, London SW7 2AZ, UK.}%

\date{\today}

\begin{abstract}

Topological waveguides supporting quantum valley Hall interfacial states confine  waves  to interfaces and, due to topological protection, are resistant to backscattering even in the presence of defects.  
These topological insulators are typically  studied by means of an infinite spectral problem. However, practical implementations are necessarily finite. In this work, we propose an alternative framework for analysing topologically non-trivial states in open, finite systems. Our approach is based on a Quasinormal Modal Expansion Method (QMEM), which directly characterizes the existence and excitation of these modes within the  open system. The resulting spectrum is complex and discrete and fully describes the topologically non-trivial states, revealing an analogy of topological mode steering as a dispersive Fabry–Pérot cavity, with a dispersion relation closely related to that of the corresponding infinite (Floquet-Bloch) ribbon problem. Our results illustrate how topologically protected waveguiding can be understood in terms of leaky cavity modes and offers a powerful framework for analysing finite topological devices. 

\end{abstract}

\maketitle



A fundamental understanding of wave energy manipulation and channeling underpins advances in electronic properties, acoustic switches, optical devices, vibration control, and filters. The ability to guide, split, and redirect waves between channels — as well as to steer them around sharp bends in a robust and lossless manner — is of significant interest across various fields of wave physics and engineering. To address these challenges, concepts from topological insulators, adapted to Newtonian wave systems, have recently generated considerable interest within the community~\cite{khanikaev2013photonic,lu2014topological,ni2023topological}. In particular, topological insulators~\cite{ozawa2019topological} have inspired devices~\cite{noh2018observation,mittal2019photonic,makwana2018designing,wiltshaw2020asymptotic,wiltshaw2023analytical,laforge2021acoustic,yan2024transmissible,li2020simulation,chaplain2020topological} that exhibit robust wave propagation, immune to certain forms of disorder and imperfections~\cite{wang2009observation,lu2014topological}.


One prominent example is the quantum valley-Hall effect (QVHE), realised in photonic and phononic crystals by breaking spatial inversion symmetry to create a non-trivial, valley-dependent band topology~\cite{pal2017edge,shalaev2019robust}. At the interface between two such crystals with inverted symmetry, valley-polarised interfacial states emerge within the bulk band gap, characterised by a well-defined valley pseudospin. These states localise energy at the interface, forming modes that exhibit negligible intervalley scattering — even at sharp corners, defects, or at the ends of the \emph{finite} valley-Hall waveguide placed in an open domain — which leads to strongly suppressed backscattering as a consequence of topological robustness.

In an \emph{infinite} valley-Hall waveguide, the edge state spans a continuous range of frequencies within the bulk band gap. However, a \emph{finite} topological waveguide of length $L$ behaves like a Fabry-Pérot resonator for the edge mode\cite{levy2017topological}. The waveguide boundaries act to quantize the allowed edge-state frequencies. Consequently, the continuous spectrum of the infinite system is replaced by a discrete set of modes in the open finite system \cite{barra1999scattering,kalozoumis2018finite}. These Fabry-Pérot–like topological modes present opportunities for compact topological devices, however we suspect topological protection is only guaranteed in the vicinity of these discrete modes. Such discrete edge resonances have been observed in photonic topological insulators\cite{muis2025broadband}.

\begin{figure*}[hbt!]
    \centering
    \includegraphics{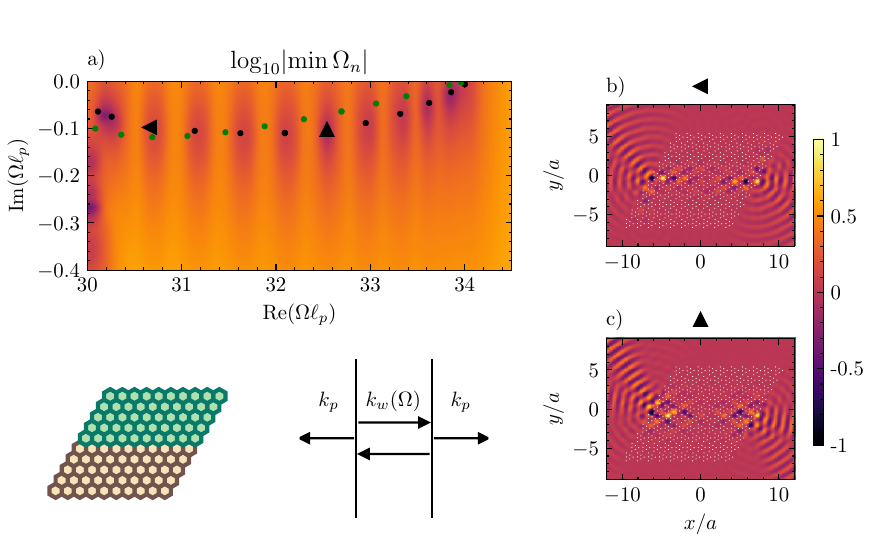}
    \caption{\label{fig:Figure_1} Spectral properties of the finite cluster. Panel \textbf{a} shows the complex spectrum for the finite cluster shown below: green and beige correspond to chiral materials. Two of the eigenstates are depicted in panels \textbf{b} and \textbf{c} (real part of the displacement field); the eigenfield is trapped at the interface between both materials, proving that they are responsible for the interface state. The diagram at the right of the cluster sketch represents the reduced-order model for the QVHE interface state, which consists of a dispersive Fabry-Perot cavity in an open space.}
\end{figure*}

While conventional theoretical approaches rely on infinite periodic structures and Floquet-Bloch theory, practical implementations are inherently finite and subject to open boundary conditions. Quasinormal modes (QNMs) provide a natural framework for this purpose, their discrete complex spectrum \cite{shestopalov1996spectral,dyatlov2019mathematical} describe both oscillation and decay. The quasinormal mode expansion method (QMEM) has been successfully applied to leaky optical cavities and resonators\cite{kristensen2014modes}, and more recently to elastic plates and metastructures \cite{vial2024quasinormal}. In the context of topological waveguides, QNMs offer a powerful tool to quantify how edge states form resonances and how topological protection exists in the vicinity of these modes.

In this Letter, we use QMEM to investigate QVHE edge states in a finite phononic crystal waveguide. Our findings reveal that the real part of the discrete complex spectrum associated with the QNMs can be viewed as a discretisation of the band structure corresponding to the infinite Floquet–Bloch ribbon problem, and offers a compelling analogy between the behaviour of QNMs and that of a dispersive Fabry–Pérot resonator. Our approach facilitates the investigation of topological physics phenomena directly within finite and open systems. Specifically, we elucidate the discrete spectrum associated with this phenomenon by mapping its characteristics onto a one-dimensional Fabry–Pérot cavity in open space, where the dispersion relation inside the cavity is determined by the eigensolution of the ribbon problem. Furthermore, we show that a superposition of QNMs accurately reconstructs the QVHE in finite lattices, offering an efficient computational approach to solving the multiple-scattering problem.


\begin{figure*}[hbt!]
    \centering
    \includegraphics{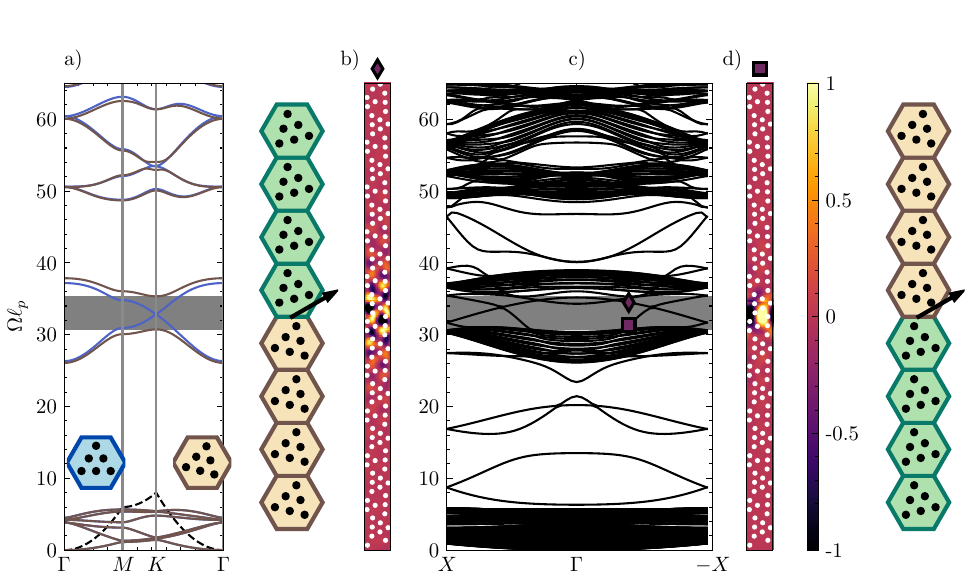}
    \caption{\label{fig:Figure_2} Spectral features of the infinite system. Panel \textbf{a} shows the band structures for the degenerate (blue) unit cell preserving the $C_{3v}$ symmetry and the broken symmetry (brown) unit cell. The gray shaded region corresponds to the topologically nontrivial bandgap. Panel \textbf{c} shows the dispersion of the ribbon problem schematic at the left; two chiral materials (green and beige) are arranged creating an interface, and the ribbon problem is solved showing the existence of two edge states in the bandgap of the doubly periodic structure, whose eigenstates are shown in panels \textbf{b} and \textbf{d}.}
\end{figure*}

We consider a two-dimensional phononic crystal composed of a hexagonal lattice of scatterers (point-like mass-spring resonators) attached to the upper surface of a thin elastic plate subjected to flexural waves (Kirchhoff-Love thin plate theory), as shown in Fig.~\ref{fig:Figure_1}. Assuming harmonic time dependence, the multiple scattering problem is addressed using a Green's function approach \cite{torrent2013elastic,martin2006multiple}, with complex resonances identified following the procedure outlined by \textit{Vial et al.} \cite{vial2024quasinormal}. The design of the underlying hexagonal cells is illustrated in Fig.~\ref{fig:Figure_2}. Fig.~\ref{fig:Figure_2}(a) demonstrates how $C_{3v}$ symmetries can produce symmetry-protected degeneracies at the $K$-point. By breaking the vertical symmetries of a cell---achieved by rotating the triangular arrangement of scatterers---this degeneracy is broken, resulting in two distinct valleys surrounding a band gap.

It is well known that such symmetry breaking~\cite{sakoda2004optical} produces a topological phase transition~\cite{makwana2018designing}
occurring between the lower and upper valleys, allowing the band gap (in gray) to be characterized as topologically non-trivial. Similarly, a topological phase transition occurs between chiral pairs of such cells. By rotating the scatterer arrangement positively or negatively to break vertical symmetries, one can create two distinct bulk materials. When these materials are joined, a valley-polarized edge mode arises at their interface within the topologically non-trivial band gap, giving rise to the quantum valley Hall effect (QVHE).

These QVHE modes are zero-line modes (ZLMs), whose formation is described in~\cite{wiltshaw2023analytical}. As expected, when stacking such chiral pairs, two ZLMs exist in the ribbon problem, as shown in Fig.~\ref{fig:Figure_2}(c). These modes are concave up and concave down, corresponding to even and odd ZLMs, respectively, and exist along topologically distinct interfaces~\cite{makwana2019tunable}.

\begin{figure}[hbt!]    \includegraphics{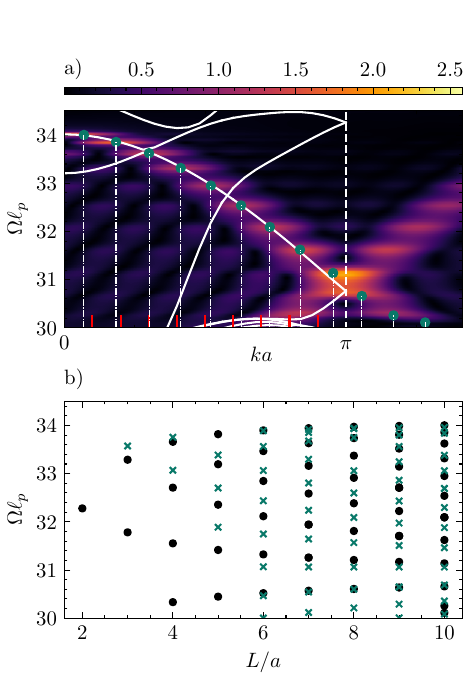}
    \caption{\label{fig:Figure_4} Fabry-Perot cavity analogue. Panel \textbf{a} shows the superposition between the ribbon dispersion, the Fourier transform of the scattering field at the interface between both materials (colormap), and the real part of the Fourier transform of the QNMs (black crosses). Panel \textbf{b} studies the evolution in the distribution of the QNMs of the QVHE waveguide as a function of the length of the waveguide. The number of solutions and their real frequency depends on the length, following the equations of a Fabry-Perot cavity.}
\end{figure}

For a finite interface of length $L$, these ZLMs become quantised into a set of allowed Fabry–Pérot resonances, determined by the band structure of the ZLM permitted to exist at the interface and the magnitude of $L$, as demonstrated in Fig.~\ref{fig:Figure_4}. Moreover, a reduced one-dimensional model can be constructed that captures only the propagation of the interface state. This is illustrated in the lower part of Fig.~\ref{fig:Figure_1}: the left panel shows the actual resonator cluster, with distinct colours indicating different chiral media and the interface aligned along the $x$-direction. The adjacent schematic presents the reduced-order model, wherein the topologically non-trivial interface — within the band gap — is effectively described as a slab of dispersive material, forming an analogue of a Fabry–Pérot cavity. The QNMs $\Omega_m$ of a Fabry–Pérot cavity satisfy the following resonance condition~\cite{muljarov2011,llorens2014absorption}

\begin{equation}
    k_w(\Omega_m) L = \pi m - i \log \frac{k_p+k_w(\Omega_m)}{k_p-k_w(\Omega_m)},
    \label{eq:QNMsSlab}
\end{equation}
with $k_w$ and $k_p$ being the wavenumbers associated to the waveguide and the bare plate respectively, $L$ is the length of the cavity and $m\in\mathbb{Z}$. The dispersive properties of the plate are given by $\Omega  = \ell_p c_p \omega_p = \ell_p c_p^2 k_p^2$, with $c_p = \sqrt{\rho h/D}$, where $\ell_p$, $\omega_p$, $\rho$, $h$ and $D$ are the lattice parameter for the hexagonal unit cell, angular frequency, density, thickness and the bending stiffness of the plate ($D = Eh^3/12(1-\nu^2)$, where E is the Young's modulus and $\nu$ is the Poisson ratio) respectively. $\Omega$ is a normalised angular frequency variable \cite{torrent2013elastic}. To estimate the disperion relation of the waveguide ($k_w(\Omega)$), it is necessary to compute the spectral properties of the corresponding infinite system.

\begin{figure*}[hbt!]
\centering    \includegraphics{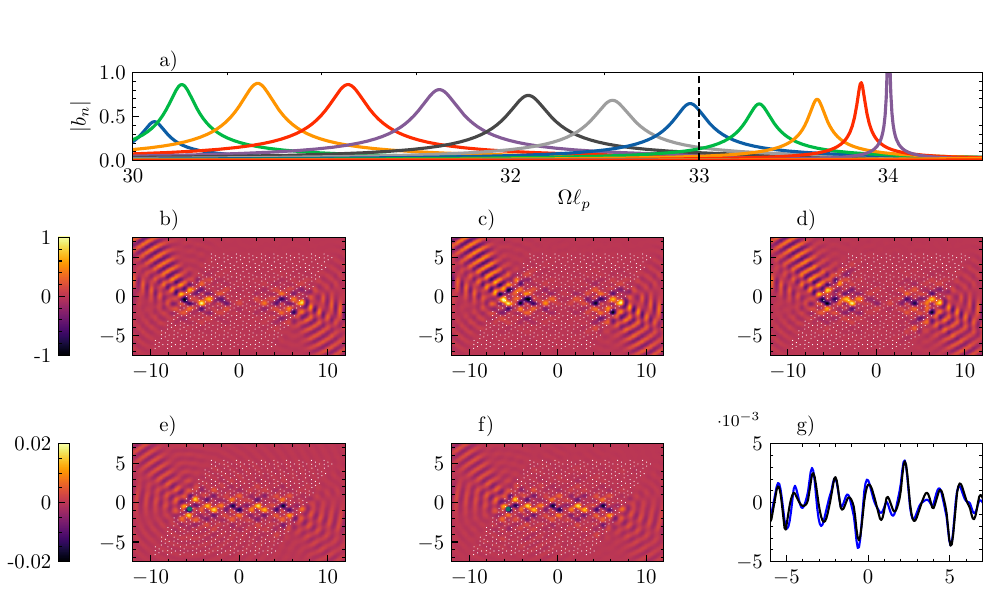}
    \caption{\label{fig:Figure_3} Scattering simulation. Panel \textbf{a} shows the evolution in frequency of the excitation coefficients. Each QNM is excited in a bell-shape manner, centered at the real frequency of the QNM and with a width inversely proportional to its Q factor. Panels \textbf{b}, \textbf{c} and \textbf{d} depict three of thee QNMs, the ones with bigger excitation coefficient for a point-like source located at the beginning of the waveguide and frequency $\Omega = 33$. These three modes are mainly the responsible for the edge state that is recovered in the scattering simulations (panels \textbf{e} and \textbf{f}). While panel \textbf{e} is a direct scattering computation solving the inverse of the multiple scattering matrix, panel \textbf{f} shows the resconstruction of the real part of the scattering field using a linear superposition of the three QNMs depicted above (panels \textbf{b}, \textbf{c} and \textbf{d}). Finally, panel \textbf{g} shows the scattering field at the interface between both chiral media, where blue represents the modal reconstruction, while black is the direct computation for the scattering field.}
\end{figure*}

The existence of the QVHE edge state is proven numerically in Fig. \ref{fig:Figure_3}. For a line source positioned at the green spot on panels \textbf{e} and \textbf{f}, the scattering field $\Phi$ is recovered using direct and modal computations. For the latter, only the QNMs found in Fig. \ref{fig:Figure_1} panel \textbf{a} have been considered. Given the good agreement between both scattering fields, we claim that the QNMs found in the previous analysis are sufficient for describing the behavior of the topologically nontrivial edge state. Panel \textbf{a} represents the moduli of the excitation coefficient for each QNM, given by the following expression 

\begin{equation}
    \Phi = \sum_n \frac{u(\Omega_n)}{u(\Omega)}\frac{\Phi_n\cdot \Psi^i}{\Omega-\Omega_n}\Phi_n = \sum_n b_n(\Omega)\Phi_n,
\end{equation}
where $\Psi^i$ is the incident field and $u(\Omega)$ is an arbitrary function, in which $u=1/k^3$ has been tested numerically to work optimally \cite{vial2024quasinormal}. The excitations are lorentzian-shaped curves centered at the real frequency of the QNM and their width is inversely proportional to the Q factor of the leaky mode.   

Within the topologically nontrivial bandgap, wave propagation is confined to the interface between the two chiral materials, as bulk propagation is prohibited. Consequently, wave transport in this frequency range can be effectively treated as a one-dimensional problem. The propagation along the waveguide is dispersive and governed by the concave down edge state, as shown in Fig. \ref{fig:Figure_2}c. At both ends of the finite cluster, we assume an interface with the unstructured (bare) plate, effectively embedding the waveguide within an open homogeneous medium.

Under this assumption, and following Eq. (\ref{eq:QNMsSlab}), the real part of the logarithmic term can be neglected, leading to an approximately equispaced distribution of wavenumbers along the waveguide. By projecting these wavenumbers $k_w$, onto the ribbon dispersion diagram, we observe that the axis is sampled at $L/a$ discrete values, where $a$ denotes the ribbon period and $L$ the total length of the waveguide. This subdivision is illustrated in Fig. \ref{fig:Figure_4}a, where red vertical segments indicate the sampled wavenumbers for a waveguide of length $L=10a$. Green points in the same panel represent the real part of the Fourier transform of the QNM eigenfields along the interface.  

The dispersion of the QNMs aligns closely with that of the concave down edge state derived from the ribbon configuration. Furthermore, the colormap in Fig.\ref{fig:Figure_4}a displays the absolute value of the Fourier transform of the scattered field along the interface, reaffirming that the observed dispersion closely follows the curve dictated by the ribbon dispersion. The wavenumbers obtained from the Fourier transform of the QNMs (green dots) do not align precisely with the red vertical segments; this discrepancy arises because the segments correspond to the ideal case of an edge state exhibiting a parabolic relation \textemdash analogous to that of a homogeneous medium.

To improve the estimation of the QNMs using the dispersive Fabry-Pérot model, the dispersion relation of the concave down edge state has been fitted and incorporated into Eq. (\ref{eq:QNMsSlab}). Specifically, the relation $k_w(\Omega) = \arccos{((\Omega-\Omega_1)/\Omega_0)}/p$ is employed, where the parameters $\Omega_0$,$\Omega_1$ and $p$ have been obtained through numerical fitting. The resulting QNMs are presented in Fig.\ref{fig:Figure_1}a as green dots in the complex frequency plane, and their real part is plotted in Fig.\ref{fig:Figure_4}d. The mean relative error between the QNMs and their estimation using the dispersive Fabry-Pérot model (black and green points in Fig. \ref{fig:Figure_1}a, respectively) is $0.44\%$.

By changing the length of the waveguide, we can change the distance and position of the QNMs of the cluster, according to equation (\ref{eq:QNMsSlab}). Figure \ref{fig:Figure_4} panel \textbf{b} shows the evolution of the real frequency of the QNMs for different waveguide lengths $L$. The length $L$ is a multiple of the length of the ribbon, so that the number of QNMs in a given finite structure will be given by the ratio between the length of the waveguide and the length of the ribbon. It is clearly seen in the picture that the number of QNMs increases with the length of the waveguide, while the frequency separation between QNMs diminishes. In the limit where $L$ tends to $\infty$, we will get QNMs acumulating in the regions where the ribbon dispersion is almost flat ($\Omega = 34$), that is to say, the density of states of the truncated system will converge to the infinite one.


\textit{In summary}, applying QNME to a finite phononic topological waveguide supporting valley-Hall edge states,  establishes a simple reduced-order model for describing the complex spectra associated with topological metamaterials. The resulting QNMs are discrete, the real part of these can be viewed as a sampling of the band structure associated with the dispersion of the infinite system (the Floquet–Bloch ribbon problem), analogous to a Fabry–Pérot cavity containing a dispersive material. These results bridge the gap between topological band theory and cavity-resonator physics, and can  guide the design of compact topological devices — such as filters, sensors, or lasers — that leverage both topological robustness and resonant field enhancement.

\begin{acknowledgments}
BV, RW and RVC are supported by the H2020 FET-proactive Metamaterial Enabled Vibration Energy Harvesting (MetaVEH) project under Grant Agreement No. 952039. MMS, SG and RVC acknowledge financial support through the DYNAMO project (101046489), funded by the European Union, but the views and opinions expressed are; however, those of the authors only and do not necessarily reflect those of the European Union or the European Innovation Council. Neither the European Union nor the granting authority can be held responsible for them.
\end{acknowledgments}

\newpage

\bibliography{apssamp}

\end{document}